\documentclass[a4paper,dvips,12pt]{article}
\usepackage{epsfig,graphics,graphicx}
\newcommand{\tV}{\widetilde{V}_x}
\newcommand{\Vef}{V_{\rm eff}}

\begin{document}

\pagestyle{myheadings}
\markright{\it P-168}
\vskip.5in
\begin{center}

%
%
\vskip.4in {\Large\bf Alternative dimensional reduction via the density
matrix: A test}
\vskip.3in
%
%
%
C.\ A.\ A.\ de Carvalho\footnote{Email: \tt aragao@if.ufrj.br} and
R.\ M.\ Cavalcanti\footnote{Email: \tt rmoritz@if.ufrj.br} \\
Instituto de F\'{\i}sica, Universidade Federal do Rio de Janeiro \\
Caixa Postal 68528, 21941-972 Rio de Janeiro, RJ
%
%
%
\end{center}
%
\vskip.2in
\begin{abstract}
We derive and analyze the perturbation series for
the classical effective potential in quantum statistical 
mechanics, treated as a toy model for the dimensionally
reduced effective action in quantum field theory at
finite temperature. The first few terms of the series
are computed for the harmonic oscillator and
the quartic potential.
\end{abstract}
%
%
%
\section{Introduction}

Recently, a procedure was devised to rewrite the density matrix
of a quantum field in thermal equilibrium at finite temperature $T$
as the exponential of an effective action in one less 
dimension \cite{CCS}.
This procedure has been termed {\em alternative dimensional reduction},
to distinguish it from the conventional dimensional reduction
that occurs at infinite $T$ \cite{weinberg}, 
the latter being a special case of the former.
The construction of such a dimensionally reduced effective action (DREA)
is in fact equivalent to the construction of
a Landau-Ginzburg ``coarse-grained free energy'' from a 
microscopic Hamiltonian, which can be further analyzed using the
powerful methods of renormalization group theory.

The construction of the DREA cannot, in general, be performed
exactly. It requires the use of perturbation theory or a
dressed-loop expansion. The purpose of this work is
to perform such a construction in the simpler context of
quantum mechanics, in order to test the accuracy of the 
truncated perturbation series for the classical effective
potential $\Vef$, the quantum mechanical version of the 
DREA.\footnote{Because we have such an objective in mind, 
we shall not discuss the semiclassical \cite{semiclass}
or variational \cite{variational} expansion of $\Vef$,
which, although generally yielding better results than
perturbation theory, are much harder to implement
in quantum field theory.} The examples
considered here are the harmonic oscillator and the quartic
potential.


\section{Density matrix and effective classical potential}

The diagonal elements of the thermal density matrix for a
particle of mass $m$ in the presence of the potential $V(x)$
can be written as a path integral \cite{feynman,kleinert}
($\hbar=m=1$):
\begin{equation}
\label{rho1}
\rho(x,x;\beta)=\int_{y(0)=x}^{y(\beta)=x}{\cal D}y\,
\exp\bigg\{-\int_0^{\beta}d\tau\left[\frac{1}{2}\,\dot{y}^2
+V(y)\right]\bigg\}.
\end{equation}
In the classical limit, the temperature is high and
$\beta$ $(=1/kT)$ is small. In this case, it is reasonable
to assume that the most important paths are those
for which $y(\tau)\approx x$.
Then Eq.\ (\ref{rho1}) can be approximated by
\begin{equation}
\label{cl_approx}
\rho(x,x;\beta)\approx e^{-\beta V(x)}\,
\rho_0(x,x;\beta),
\end{equation}
where \cite{feynman,kleinert}
\begin{equation}
\label{rho0}
\rho_0(x,x;\beta)=\int_{y(0)=x}^{y(\beta)=x}{\cal D}\eta\,
\exp\left\{-\int_0^{\beta}\frac{1}{2}\,\dot{\eta}^2\,d\tau\right\}
=\frac{1}{\sqrt{2\pi\beta}}
\end{equation}
is the diagonal element of the density matrix for the free 
particle. Inserting
Eq.\ (\ref{rho0}) into Eq.\ (\ref{cl_approx}) and integrating
over $x$ one obtains the classical partition function,
\begin{equation}
Z_{\rm cl}=\int_{-\infty}^{\infty}
\frac{dx}{\sqrt{2\pi\beta}}\,e^{-\beta V(x)}.
\end{equation}
%

In order to improve the approximation (\ref{cl_approx}) 
in a systematic way, we shall put 
$y(\tau)=x+\eta(\tau)$, with $\eta(0)=\eta(\beta)=0$,
into Eq.\ (\ref{rho1}) and compute the path integral
perturbatively. Thus,
\begin{equation}
\rho(x,x;\beta)=e^{-\beta V(x)}
\int_{\eta(0)=0}^{\eta(\beta)=0}{\cal D}\eta\,
\exp\bigg\{-\int_0^{\beta}d\tau\left[\frac{1}{2}\,\dot{\eta}^2
+\tV(\eta)\right]\bigg\},
\end{equation}
where $\tV(\eta)\equiv V(x+\eta)-V(x)$. Expanding
the exponential in powers of $\tV(\eta)$, we obtain
\begin{eqnarray}
\rho(x,x;\beta)&=&e^{-\beta 
V(x)}\,\rho_0(0,0;\beta)\left\{1-\int_0^{\beta}d\tau\,\langle\tV[\eta(\tau)]\rangle_0\right.
\nonumber \\
& &+\left.\frac{1}{2!}\int_0^{\beta}d\tau_1\int_0^{\beta}d\tau_2\,
\langle\tV[\eta(\tau_1)]\,\tV[\eta(\tau_2)]\rangle_0+\cdots\right\},
\label{rho_p}
\end{eqnarray}
where 
\begin{equation}
\langle{\cal O}\rangle_0\equiv\rho_0(0,0;\beta)^{-1}
\int_{\eta(0)=0}^{\eta(\beta)=0}{\cal D}\eta\,\,{\cal O}\,
\exp\left\{-\int_0^{\beta}\frac{1}{2}\,\dot{\eta}^2\,d\tau
\right\}.
\end{equation}

If $\tV(\eta)$ is a polynomial in $\eta$, one can compute
the expectation values in Eq.\ (\ref{rho_p}) with the help
of Wick's theorem:
\begin{equation}
\label{Wick}
\langle\eta(\tau_1)\cdots\eta(\tau_n)\rangle_0=
\sum_P G_0(\tau_{P(1)},\tau_{P(2)})\cdots
G_0(\tau_{P(n-1)},\tau_{P(n)}) 
\end{equation}
if $n$ is even, and zero otherwise. The sum $\sum_P$ runs over
all permutations that lead to different 
expressions,\footnote{Two permutations $P$ and $P'$ lead to
the same expression if the sets $\{\{P(1),P(2)\},\ldots,
\{P(n-1),P(n)\}\}$ and $\{\{P'(1),P'(2)\},\ldots,
\{P'(n-1),P'(n)\}\}$ are equal.}
and $G_0(\tau,\tau')=\langle\eta(\tau)\eta(\tau')\rangle_0$.
It satisfies 
\begin{equation}
\label{G}
-\frac{\partial^2}{\partial\tau^2}\,G_0(\tau,\tau')
=\delta(\tau-\tau'),\qquad G_0(0,\tau')=G_0(\beta,\tau')=0.
\end{equation} 
Eq.\ (\ref{G}) can be easily solved, yielding
\begin{equation}
\label{G_0}
G_0(\tau,\tau')=\frac{(\beta-\tau')\,\tau}{\beta}\,
\theta(\tau'-\tau)+\frac{(\beta-\tau)\,\tau'}{\beta}\,\theta(\tau-\tau'),
\end{equation}
where $\theta(\tau)$ is the Heaviside step function.

As a final step, we shall resum the terms in brackets
in Eq.\ (\ref{rho_p}) into an exponential. The result 
can be cast as
\begin{equation}
\label{rho_eff}
\rho(x,x;\beta)=\frac{1}{\sqrt{2\pi\beta}}\,\exp\left\{
-\beta \Vef(x)\right\},
\end{equation}
where
\begin{equation}
\label{V_eff}
\beta \Vef(x)=\beta V(x)
+\int_0^{\beta}d\tau\,\langle\tV[\eta(\tau)]\rangle_{0,c}
-\frac{1}{2!}\int_0^{\beta}d\tau_1\int_0^{\beta}d\tau_2\,
\langle\tV[\eta(\tau_1)]\,\tV[\eta(\tau_2)]\rangle_{0,c}
+\cdots
\end{equation}
The index $c$ in the expectation values
means that only connected diagrams, or cumulants,
should be taken into account when applying Wick's 
theorem.\footnote{For instance,
$\langle\eta^2(\tau_1)\,\eta^2(\tau_2)\rangle_{0}=
2\,G_0^2(\tau_1,\tau_2)+G_0(\tau_1,\tau_1)\,G_0(\tau_2,\tau_2)$,
whereas
$\langle\eta^2(\tau_1)\,\eta^2(\tau_2)\rangle_{0,c}=
2\,G_0^2(\tau_1,\tau_2)$.}

Equation (\ref{V_eff}) formally defines the {\em effective
classical potential} $\Vef(x)$ as a power series in $x$,
with coefficients that are functions of the temperature.
It is the quantum mechanical analogue of the Landau-Ginzburg
functional in statistical mechanics.
In the next section we shall compute the first few terms
of the cumulant expansion of $\Vef(x)$ for the harmonic
oscillator and the quartic potential.


\section{Examples}

\subsection{Harmonic oscillator}

As our first example, let us consider the harmonic oscillator:
\begin{equation}
V(x)=\frac{1}{2}\,\omega^2x^2,\qquad
\tV(\eta)=\frac{1}{2}\,\omega^2(\eta^2+2x\eta).
\end{equation}
The first two cumulants give us
\begin{eqnarray}
\langle\tV[\eta(\tau)]\rangle_{0,c}&=&
\frac{1}{2}\,\omega^2G_0(\tau,\tau),
\\
\langle\tV[\eta(\tau_1)]\,\tV[\eta(\tau_2)]\rangle_{0,c}
&=&\frac{1}{4}\,\omega^2\left[\langle\eta^2(\tau_1)\eta^2(\tau_2)
\rangle_{0,c}+4x^2\langle\eta(\tau_1)\eta(\tau_2)
\rangle_{0,c}\right]
\nonumber \\
&=&\frac{1}{4}\,\omega^2\left[2G_0^2(\tau_1,\tau_2)
+4x^2G_0(\tau_1,\tau_2)\right].
\end{eqnarray}
Inserting these results into Eq.\ (\ref{V_eff}) and
computing the integrals we obtain
\begin{equation}
\label{osc_2}
\beta \Vef(x)=\left(\frac{1}{12}\,\beta^2\omega^2-\frac{1}{360}\,
\beta^4\omega^4+\cdots\right)+\omega x^2\left(\frac{1}{2}\,\beta\omega
-\frac{1}{24}\,\beta^3\omega^3+\cdots\right).
\end{equation}
This result should be compared with the exact result
for the harmonic oscillator,\footnote{Eq.\ (\ref{osc_ex})
is obtained by expressing $\rho(x,x;\beta)$
for the harmonic oscillator (see Refs.\ \cite{feynman,kleinert}) 
in the form of Eq.\ (\ref{rho_eff}).}
\begin{equation}
\beta V_{\rm eff,ex}(x)=\frac{1}{2}\,
\ln\left(\frac{\sinh\beta\omega}{\beta\omega}\right)
+\omega x^2\tanh\frac{\beta\omega}{2}.
\label{osc_ex}
\end{equation}
One can easily check that Eq.\ (\ref{osc_2}) correctly reproduces
the first few terms of the expansion of $\beta V_{\rm eff,ex}(x)$ in powers of 
$\beta$.


\subsection{Quartic potential}

Let us now consider the quartic potential:
\begin{equation}
\label{quartic}
V(x)=\frac{\lambda}{4}\,x^4,\qquad\tV(\eta)=
\frac{\lambda}{4}\,(\eta^4+4x\eta^3+6x^2\eta^2+4x^3\eta).
\end{equation}
In this case, the expansion of the effective classical potential
in cumulants, Eq.\ (\ref{V_eff}), is also an expansion in
powers of $\lambda$. Thus, to second order in $\lambda$, we have
\begin{eqnarray}
\Vef(x)&=&V(x)+\frac{\lambda}{4\beta}\int_0^{\beta}d\tau
\left[\langle\eta^4(\tau)\rangle_{0,c}
+6x^2\langle\eta^2(\tau)\rangle_{0,c}\right]+O(\lambda^2)
\nonumber \\
&=&\frac{\lambda}{4}\,x^4+
\frac{\lambda}{4\beta}\int_0^{\beta}d\tau\left[
3G_0^2(\tau,\tau)+6x^2G_0(\tau,\tau)\right]+O(\lambda^2)
\nonumber \\
&=&\frac{\lambda}{4}\,x^4+\frac{\lambda}{4}\,\beta x^2
+\frac{\lambda}{40}\,\beta^2+O(\lambda^2).
\label{V1}
\end{eqnarray}
The second term on the r.h.s.\ of Eq.\ (\ref{V1}) is analogous
to the radiatively induced thermal mass in the massless 
$\lambda\varphi^4$ model in field theory at finite 
temperature \cite{kapusta}.

There occurs a problem when one computes the $O(\lambda^2)$
correction to $\Vef(x)$. It is a sixth
degree polynomial in $x$, with the coefficient of $x^6$
given by
\begin{equation}
-\frac{\lambda^2}{2!\,\beta}\int_0^{\beta}d\tau_1
\int_0^{\beta}d\tau_2\,\langle\eta(\tau_1)\eta(\tau_2)\rangle_{0,c}
=-\frac{\lambda^2}{24}\,\beta^2<0.
\end{equation}
It follows that if one discards cubic and higher order
terms in $\lambda$ in the expansion (\ref{V_eff}), one
ends up with a classical effective potential which is
unbounded from below, with the disastrous consequence 
that $\rho(x,x;\beta)$ diverges
as $|x|\to\infty$. One can remedy this problem by going
to the next order in the expansion (\ref{V_eff}).
At this level of approximation, $\Vef(x)$ is 
an eighth degree polynomial in $x$, the coefficient
of the highest power of $x$ now being a positive number,
given by
\begin{equation}
\frac{\lambda^3}{3!\,\beta}\cdot\frac{96}{64}
\int_0^{\beta}d\tau_1\cdots\int_0^{\beta}d\tau_3\left[
\langle\eta^2(\tau_1)\eta(\tau_2)\eta(\tau_3)\rangle_{0,c}
+\mbox{cyclic perm.}\right]=\frac{\lambda^3}{80}\,\beta^4.
\end{equation}

Instead of expanding $\Vef(x)$ in powers of $\lambda$, one
may expand it in powers of $\beta$ --- a {\em high temperature
expansion}. Using Wick's theorem and
the explicit form of $G_0(\tau,\tau')$, 
and making the change of variables $\tau_j=\beta\tau_j'$ in the
integral below, one can easily show that
\begin{equation}
\frac{1}{\beta}\int_0^{\beta}d\tau_1\cdots\int_0^{\beta}d\tau_n\,
\langle\eta^{k_1}(\tau_1)\cdots\eta^{k_n}(\tau_n)\rangle_{0,c}
\propto\beta^{n-1+(k_1+\cdots+k_n)/2}.
\end{equation}
In terms of Feynman diagrams, this means that a (connected) graph 
with $V$ vertices and $L$ lines gives a contribution to
$\Vef(x)$ proportional to $\beta^{V+L-1}$. 
Thus, we have
\begin{eqnarray}
\Vef(x)&=&\frac{\lambda}{4}\,x^4+\beta\,\frac{\lambda}{4}\,x^2
+\beta^2\left(\frac{\lambda}{40}-\frac{\lambda^2}{24}\,x^6\right)
-\beta^3\,\frac{3\lambda^2}{40}\,x^4
\nonumber \\
& &+\beta^4\left(-\frac{71\lambda^2}{3360}\,x^2
+\frac{\lambda^3}{80}\,x^8\right)+O(\beta^5).
\end{eqnarray} 
Notice that the second and third order approximations to
$\Vef(x)$ are unbounded from below, hence physically
nonsensical. Therefore,
like the expansion in powers of $\lambda$, the expansion
in powers of $\beta$ cannot be truncated at an arbitrary
order, but only at those orders for which the
resulting approximation to $\Vef(x)$ is bounded from below.


\section{Conclusions}

The perturbative construction of the DREA in quantum mechanics is, therefore, 
equivalent
to a perturbative expansion of the classical effective potential in powers of 
$x$. The construction yields a high-temperature series in $\beta$, or a series 
in the coupling constant $\lambda$, as shown in the quartic case.

The generalization to field theories \cite{CCS} will also yield perturbative 
expansions in powers of the ``boundary'' fields, i.e., the field 
configurations on the euclidean time boundaries at $\tau=0$ and $\tau=\beta$, 
a role played by $x$ in quantum mechanics. The coefficients will, in general, 
depend on temperature, couplings, and on an ultraviolet cutoff.
The boundary fields are independent of the euclidean time $\tau$ (they only 
depend on spatial coordinates). Thus, their effective action accomplishes the 
dimensional reduction advertised.
For practical applications, a truncation of the series will be required, just 
as in quantum mechanics.

There is, however, an additional complication in field theory: the terms in 
the effective action are, in general, nonlocal in the boundary fields. The 
coefficient of the term involving $n$ boundary fields is given by an $n$-point 
connected Green function of the fluctuations around a ``background'' field. 
The latter is defined as the solution to the free field equation of motion 
that interpolates between the boundary field at $\tau=0$ and $\tau=\beta$ (in 
our quantum mechanical examples, the background field coincides with the 
boundary field, both being given by $x$; in general, the background is 
$\tau$-dependent).

Hopefully, both difficulties --- the need to truncate and the nonlocality --- 
can be circumvented in certain cases. Renormalization theory will dictate the 
dependence of the coefficients on the ultraviolet cutoff, so that 
nonrenormalizable terms will probably not contribute in the continuum limit, 
leading to a natural truncation. If, in addition, we investigate the infrared 
limit, as in the study of critical behavior near phase transitions, we might 
restrict our attention to the zero momentum limit of the coefficients, and 
recover a local structure. In summary, we hope that the combination of 
ultraviolet and infrared limits will restrict the form of the effective 
action, and give us a Landau-Ginzburg free energy of the expected form, whose 
parameters we will be able to compute from the underlying microscopic theory.


\section*{Acknowledgments}

R.M.C.\ acknowledges the financial support from CNPq. 
C.A.A.C.\ acknowledges support from
CNPq and FUJB/UFRJ. 

%
%
%
%
%

%

\begin{thebibliography}{99}

\bibitem{CCS} C. A. A. de Carvalho, J. M. Cornwall,
and A. J. da Silva, Phys. Rev. D {\bf 64}, 025021 (2001).

\bibitem{weinberg} S.\ Weinberg, 
Phys. Lett. {\bf 91B}, 51 (1980); 
P. Ginzparg, Nucl. Phys. {\bf B170}, 388 (1980); 
T. Applequist and R. D. Pisarski, 
Phys. Rev. D {\bf 23}, 2305 (1981).

\bibitem{semiclass} C. DeWitt-Morette, 
Commun. Math. Phys. {\bf 28}, 47 (1972); {\bf 37}, 63 (1974);
Ann. Phys. (N.Y.) {\bf 97}, 367 (1976);
M. M. Mizrahi, J. Math. Phys. {\bf 17}, 566 (1976);
{\bf 19}, 298 (1978); {\bf 20}, 844 (1979);
M. Roncadelli, Phys. Rev. Lett. {\bf 72}, 1145 (1994);
C. A. A. de Carvalho, R. M. Cavalcanti,
E. S. Fraga, and S. E. Jor\'as, 
Ann. Phys. (N.Y.) {\bf 273}, 146 (1999).

\bibitem{variational} R. Giachetti and V. Tognetti,
Phys. Rev. Lett. {\bf 55}, 912 (1985);
R. P. Feynman and H. Kleinert,
Phys. Rev. A {\bf 34}, 5080 (1986);
A. Cuccoli {\it et al.}, 
J. Phys.: Condens. Matter {\bf 7}, 7891 (1995);
H. Kleinert, W. K\"uzinger, and A. Pelster,
J. Phys. A: Math. Gen. {\bf 31}, 8307 (1998).

\bibitem{feynman} R. P. Feynman, {\it Statistical Mechanics:
A Set of Lectures}
(W. A. Benjamin, Reading, Massachusetts, 1972).

\bibitem{kleinert} H. Kleinert, {\it Path Integrals in
Quantum Mechanics, Statistics, and Polymer Physics}
(World Scientific, Singapore, 1995).

\bibitem{kapusta} J. I. Kapusta, {\it Finite Temperature
Field Theory} (Cambridge University Press, Cambridge,
England, 1989); M. Le Bellac, {\it Thermal Field Theory}
(Cambridge University Press, Cambridge,
England, 1996).

\end{thebibliography}
\end{document}